\documentclass[twocolumn,prl,preprintnumbers,superscriptaddress,amsmath,amssymb,english]{revtex4}
\usepackage{amsmath,amssymb,amsfonts,mathrsfs}
\usepackage{amsthm}
\usepackage{CJK}
\usepackage{bm}
\usepackage{babel}
\usepackage{graphicx}
\allowdisplaybreaks
\usepackage{braket}
\usepackage[breaklinks]{hyperref}
\usepackage{color}
\usepackage{epstopdf}
\hypersetup{colorlinks=true, linkcolor=blue, citecolor=blue, filecolor=blue, urlcolor=blue}

\begin{document}

\title{Robustness of Half-Integer Quantized Hall Conductivity against Disorder in an Anisotropic Dirac Semimetal with Parity Anomaly}

\author{Zhen Ning}
\affiliation{Institute for Structure and Function $\&$ Department of Physics $\&$ Chongqing Key Laboratory for Strongly Coupled Physics, Chongqing University, Chongqing 400044, People's Republic of China}
\affiliation{Center of Quantum Materials and Devices, Chongqing University, Chongqing 400044, People's Republic of China}

\author{Xianyong Ding}
\affiliation{Institute for Structure and Function $\&$ Department of Physics $\&$ Chongqing Key Laboratory for Strongly Coupled Physics, Chongqing University, Chongqing 400044, People's Republic of China}
\affiliation{Center of Quantum Materials and Devices, Chongqing University, Chongqing 400044, People's Republic of China}

\author{Dong-Hui Xu}
\email[]{donghuixu@cqu.edu.cn}
\affiliation{Institute for Structure and Function $\&$ Department of Physics $\&$ Chongqing Key Laboratory for Strongly Coupled Physics, Chongqing University, Chongqing 400044, People's Republic of China}
\affiliation{Center of Quantum Materials and Devices, Chongqing University, Chongqing 400044, People's Republic of China}

\author{Rui Wang}
\email[]{rcwang@cqu.edu.cn}
\affiliation{Institute for Structure and Function $\&$ Department of Physics $\&$ Chongqing Key Laboratory for Strongly Coupled Physics, Chongqing University, Chongqing 400044, People's Republic of China}
\affiliation{Center of Quantum Materials and Devices, Chongqing University, Chongqing 400044, People's Republic of China}

\date{\today}

\begin{abstract}
Two-dimensional Dirac semimetals with a single massless Dirac cone exhibit the parity anomaly. Usually, such a kind of anomalous topological semimetallic phase in real materials is unstable where any amount of disorder can drive it into a diffusive metal and destroy the half-integer quantized Hall conductivity as an indicator of parity anomaly. Here, based on low-energy effective model, we propose an anisotropic Dirac semimetal which explicitly breaks time-reversal symmetry and carries a half-integer quantized Hall conductivity. This topological semimetallic phase can be realized on a deformed honeycomb lattice subjected to a magnetic flux. Moreover, we perceptively investigate the disorder correction to the Hall conductivity. The results show that the effects of disorder can be strongly suppressed and thereby the nearly half-integer quantization of Hall conductivity can exist in a wide region of disorder, indicating that our proposed anisotropic Dirac semimetal is an exciting platform to investigate the parity anomaly phenomena.
\end{abstract}
\maketitle

\emph{{\color{magenta}Introduction.}}---The discovery of Dirac matter is one of the particularly impressive achievements of modern condensed matter physics. Dirac materials, in which the low-energy electronic excitations are governed by the Dirac equation, provide a natural playground for exploring exotic phenomena in the relativistic quantum field theory. A prominent example is
the parity anomaly, which occurs in two-dimensional (2D) Dirac semimetals with a single Dirac cone~\cite{PA1,PA2,PA3}. In this case, the Dirac fermions in $(2+1)$ space-time dimensions interacting with a fluctuating gauge field lead to the breaking of
time-reversal symmetry or reflection symmetry~\cite{PA1,PA2,PA3,PA4,PA5,witten2016prb,WilsonF}.
A remarkable consequence of the parity anomaly is the half-integer quantized Hall conductivity as predicted by the anomaly-induced Chern-Simons theory~\cite{Haldane1988,PA-CM1,PA-CM2,Ludwig1994,halfQAH1,halfQAH2}.
In recent years, the parity anomaly has been intensely investigated in the context of topological insulators (TIs) and even artificial periodic structures~\cite{bhzPA,PA-exp,QAH,MTI1,MTI2,MTI3,realz1,realz2,realz3,realz4,realz5}.
On the other hand, the no-go theorem of fermion doubling forces Dirac fermions to come in pairs, which hinders the experimental detection of parity anomaly phenomena. Very recently, the evidence of parity anomaly with half-integer quantized Hall conductivity was observed in semimagnetic TIs~\cite{PA-exp1,PA-exp2} where the massless Dirac fermions is only present at one surface while the Dirac state at opposite surface is massive.
These experimental observations demonstrated that the single-node Dirac semimetal provides a promising platform to realize the parity anomaly in condensed matter physics~\cite{bhzQASM}. 

The stability of Dirac semimetals in the presence of disorder is a long-sought question of fundamental importance. The disorder usually destroys the Dirac semimetallic states and thereby induces a finite density of states at the Fermi level \cite{CastroNeto2009,DasSarma,Pixley2021,Syzranov2018}.
In 3D Dirac 
semimetals, the analysis of renormalization group shows the weak disorder is a irrelevant perturbation, and thus these systems remain stable up to a critical strength of disorder~\cite{Fradkin1986,Goswami2011,Pixley2015}. However, the disorder is (marginally) relevant in 2D Dirac semimetals, in which any finite amount of disorder can generate a low energy scale $E_c$, blow which the density of states approaches to a finite value \cite{CastroNeto2009,PatrickLee,Nersesyan,Ando,Efetov2006,Altland,Mirlin2006,Ning2020}. In this regard, the 2D Dirac semimetals are unstable against disorder and will evolve into diffusive metals. Therefore, the unique phenomenon of parity anomaly can always be smeared in the presence of disorder. To investigate various intrinsic topological effects derived from the parity anomaly, it is desirable to suppress the influence of disorder and stabilize the Dirac semimetallic states with half-integer quantized Hall conductivity in 2D systems.

In this work, we first propose a 2D topological Dirac semimetal which avoids fermion doubling problem and thus only hosts a single Dirac cone of massless fermions.
Specifically, as a signature of parity anomaly, the intrinsic half-integer quantized Hall conductivity $\sigma^\text{int}_{xy} = \frac{e^2}{2h}$ is directly obtained in the absence of disorder. We then construct a lattice model, which explicitly breaks time-reversal symmetry, to realize this topological Dirac semimetal with parity anomaly. The presence of disorder imposes a correction $\delta\sigma_{xy}$, which is proportional to a damping parameter $\eta$, to the intrinsic Hall conductivity.
Using the Kubo-Streda formula, we derive an explicit expression for the correction $\delta\sigma_{xy}$, which guides us how to preserve
the half-integer quantized Hall conductivity. We observe a crossover from a diffusive metal to the topological Dirac semimetal. Remarkably, we find that the nearly half-integer quantization of Hall conductivity can exist in a wide region of disorder and in a relatively wide energy window around the Fermi level. 

\emph{{\color{magenta}Realization of 2D Dirac semimetals with the parity anomaly.}}--- First of all, we propose that the parity anomaly can be realized in a 2D topological semimetal with a single anisotropic Dirac cone. The low-energy effective Hamiltonian relative to the Dirac point $\mathbf{D}$ reads
\begin{equation}
\begin{aligned}
\mathrm{h}(\mathbf{k}) &= \mathrm{d}_x({\mathbf{k}})\sigma_x + \mathrm{d}_y({\mathbf{k}})\sigma_y +\mathrm{d}_z({\mathbf{k}})\sigma_z\\
&= (c_xk^2_x + c_yk^2_y)\sigma_x + vk_y\sigma_y +\lambda k_x\sigma_z,
\end{aligned}
\label{eq:hk}
\end{equation}
where $\boldsymbol{\sigma}=(\sigma_x,\sigma_y,\sigma_z)$ is the Pauli matrixes acting on the sublattice space, and model parameters $c_x$, $c_y$, $v$, and $\lambda$ depend on the realistic lattices. It is worth noting that the spin degrees of freedom trivially double the Hilbert space and thus we here only focus on one spin sector \cite{CPLgra}. In the absence of $\lambda k_x\sigma_z$ term, the effective Hamiltonian Eq.~(\ref{eq:hk}) returns to the usual semi-Dirac model \cite{semi-Dirac-1, semi-Dirac-2} which displays linear dispersion along one direction and quadratic dispersion along the other. Moreover, the presence of $\lambda k_x\sigma_z$ term breaks time-reversal symmetry $\mathcal{T}$ and thus can lead to distinct topological phases. To reveal the topological features, we calculate the anomalous Hall conductivity using Berry curvature as
\begin{equation}
	\begin{aligned}
		\Omega(\mathbf{k}) = -\frac{\mathbf{d}_{\mathbf{k}}\cdot\partial_{k_x}\mathbf{d}_{\mathbf{k}}\times\partial_{k_y}\mathbf{d}_{\mathbf{k}}}{2\mathrm{d}^3_{\mathbf{k}}}
		= -\frac{v\lambda \mathrm{d}_x(\mathbf{k})}{2\mathrm{d}^3_{\mathbf{k}}}
	\end{aligned}
	\label{eq:BC}
\end{equation}
where the vector $\mathbf{d}_{\mathbf{k}}=(\mathrm{d}_x,\mathrm{d}_y,\mathrm{d}_z)$ is given by Eq.~(\ref{eq:hk}). 
 Then, we obtain the intrinsic Hall conductivity as
\begin{equation}
	\begin{aligned}
		\sigma_{xy} = \frac{e^2}{\hbar}\int \frac{d^2\mathbf{k}}{(2\pi)^2} \Omega(\mathbf{k}) = -\frac{e^2}{2h}\mathrm{sgn}(v)\mathrm{sgn}(\lambda),
	\end{aligned}
	\label{eq:AHC}
\end{equation}
where we assume parameters $c_{x}$ and $c_{y}$ in Eq.~(\ref{eq:hk}) are positive. As shown in Eq.~(\ref{eq:AHC}), it is found that the $\lambda k_x\sigma_z$ term dominates the emergence of half-integer quantized Hall conductivity and thus brings a distinct topological phase.

\begin{figure}
	\centering
	\includegraphics[width=0.47\textwidth]{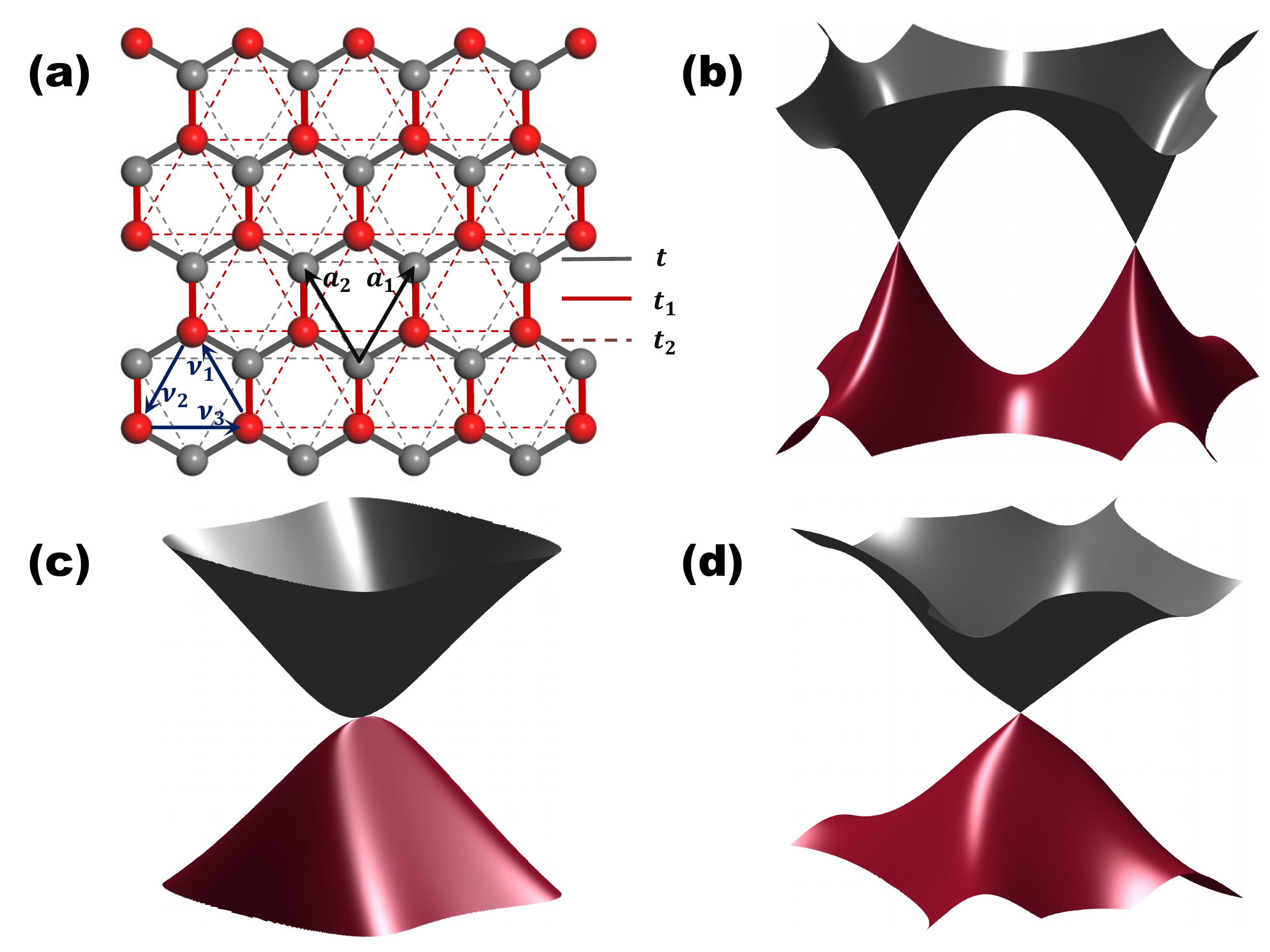}%
	\caption{ The evolution of the band structure of deformed honeycomb lattice. (a) The deformed honeycomb model. The primitive vectors of Bravais lattice are $\mathbf{a}_{1} = (-\sqrt{3}/2,3/2)$, $\mathbf{a}_{1} = (\sqrt{3}/2,3/2)$. The vectors of next-nearest neighbors are $\boldsymbol{\nu}_{1} = (\sqrt{3},0)$, $\boldsymbol{\nu}_{2} = (-\sqrt{3}/2,3/2)$,$\boldsymbol{\nu}_{3} = (-\sqrt{3}/2,-3/2)$. The (anisotropic) nearest-neighbor hopping amplitude $t (t_1)$ and the next-nearest-neighbor hopping ampitude $t_2$ are also denoted.
(b) Two Dirac cones in Brillouin zone for the isotropic honeycomb case (i.e., $t =t_1$). (c) The semi-Dirac semimetal arising from the merging of two Dirac nodes as $t_1 \rightarrow 2t$ with $t_2=0$. (d) The topological Dirac semimetal with single Dirac cone occurs when $t_1 =2t$ and $t_2\neq0$.}
\label{Fig:Figure}
\end{figure}

The continuum model Eq.~(\ref{eq:hk}) can be realized on the deformed honeycomb lattice as depicted in Fig.~\ref{Fig:Figure}(a), and the corresponding tight-binding Hamiltonian with anisotropic hopping parameters can be written as $\hat{\mathrm{H}}_t = \sum_{nn}t_{ij}c^{\dagger}_ic_j + \sum_{nnn}t_2 e^{i\phi_{ij}}c^{\dagger}_ic_j+h.c.$, where $t_{ij}$ are the nearest-neighbor (NN) hopping energies $t (t_1)$ and  $t_2 e^{i\phi_{ij}}$ is the next-nearest-neighbor (NNN) hopping with staggered magnetic flux. The lattice model is similar to the Haldane's model~\cite{Haldane1988}, and we also propose a possible  realization of such model Hamiltonian as discussed in the Supplemental Material~\cite{SM}. In the reciprocal space, the Hamiltonian can be represented by the two component Bloch states $\psi^{\dagger}_{\mathbf{k}} = (c^{\dagger}_{A\mathbf{k}},c^{\dagger}_{B\mathbf{k}})$ as,
\begin{equation}
\begin{aligned}
\hat{\mathrm{H}}_t(\mathbf{k}) = \sum_{\mathbf{k}} \psi^{\dagger}_{\mathbf{k}}
\begin{pmatrix} \Delta_{\mathbf{k}} & f_{\mathbf{k}} \\ f^{*}_{\mathbf{k}} & -\Delta_{\mathbf{k}} \end{pmatrix}\psi_{\mathbf{k}},
\end{aligned}
\label{eq:ht}
\end{equation}
where we define $f_{\mathbf{k}} = t_1 + te^{-i\mathbf{k}\cdot \mathbf{a}_1} + te^{-i\mathbf{k}\cdot \mathbf{a}_2}$ and $\Delta_{\mathbf{k}} = -t_2/\sqrt{3}\sum_{i}\sin(\mathbf{k}\cdot \boldsymbol{\nu}_i)$. The primitive vectors of Bravais lattice $\mathbf{a}_{1}, \mathbf{a}_{2}$ and the vectors of NN neighbors $\boldsymbol{\nu}_{1} ,\boldsymbol{\nu}_{2},\boldsymbol{\nu}_{2} $ are shown in Fig.~\ref{Fig:Figure}(a). In the absence of NNN hopping term $t_2$, the two-band Hamiltonian Eq.~(\ref{eq:ht}) returns to the usual model of graphene with two Dirac points respectively located at $\mathbf{K} = (\frac{2\pi}{3\sqrt{3}a},\frac{2\pi}{3a})$ and $\mathbf{K}' = (-\frac{2\pi}{3\sqrt{3}a},\frac{2\pi}{3a})$ in Brillouin zone when $t_1 = t$ [see Fig.~\ref{Fig:Figure}(b)]. When deforming the honeycomb lattice as $t_1 \rightarrow 2t$, the two Dirac nodes will merge into a single Dirac point at $\mathbf{D} = (0,\frac{2\pi}{3a})$ at the transition point $t_1 = 2t$. Intriguingly, in sharp contrast to the usual semi-Dirac semimetal with anisotropic dispersion [see Fig.~\ref{Fig:Figure}(c)], this deformed honeycomb lattice model gives a distinct Dirac cone as shown in Fig.~\ref{Fig:Figure}(d). To illustrate this, we expand the function $f_{\mathbf{k}}$ and $\Delta_{\mathbf{k}}$ near the Dirac point $\mathbf{D}$ and obtain the low-energy effective Hamiltonian as
\begin{equation}
\begin{aligned}
\mathrm{h}_t(\mathbf{k}) &= (\delta + c_xk^2_x + c_yk^2_y)\sigma_x + vk_y\sigma_y +\lambda k_x\sigma_z,
\end{aligned}
\label{eq:hking}
\end{equation}
where the wave vector $\mathbf{k} = (k_x,k_y)$ is relative to the Dirac point $\mathbf{D}$. The parameters in Eq.~(\ref{eq:hking}) are given by
$\delta =-2t+t_1$, $c_x = \frac{3ta^2}{4}$, $c_y = \frac{9ta^2}{4}$, and the Fermi velocities along the $x$ and $y$ directions are defined as $\lambda = 4t_2a$, $v = -3ta$. We note that the NNN hopping $t_2$ is responsible for the presence of topological phase with half-integer quantized Hall conductivity. Using the formula Eqs.~(\ref{eq:BC}) and (\ref{eq:AHC}), the Hall conductivity can be expressed as $\sigma_{xy} = \frac{e^2}{2h}[1-\mathrm{sgn}(\delta)]$. Therefore, the effective Hamiltonian Eq.~(\ref{eq:hking}) describes the transition from normal insulator when $\delta>0$ into a Chern insulator when $\delta<0$. At the transition point $\delta=0$ (i.e., $t_1 = 2t$), we obtain a topological semimetal carrying a half-integer quantized Hall conductivity $\sigma_{xy}= \frac{e^2}{2h}$. 

\begin{figure}[htb]
	\centering
	\includegraphics[width=0.47\textwidth]{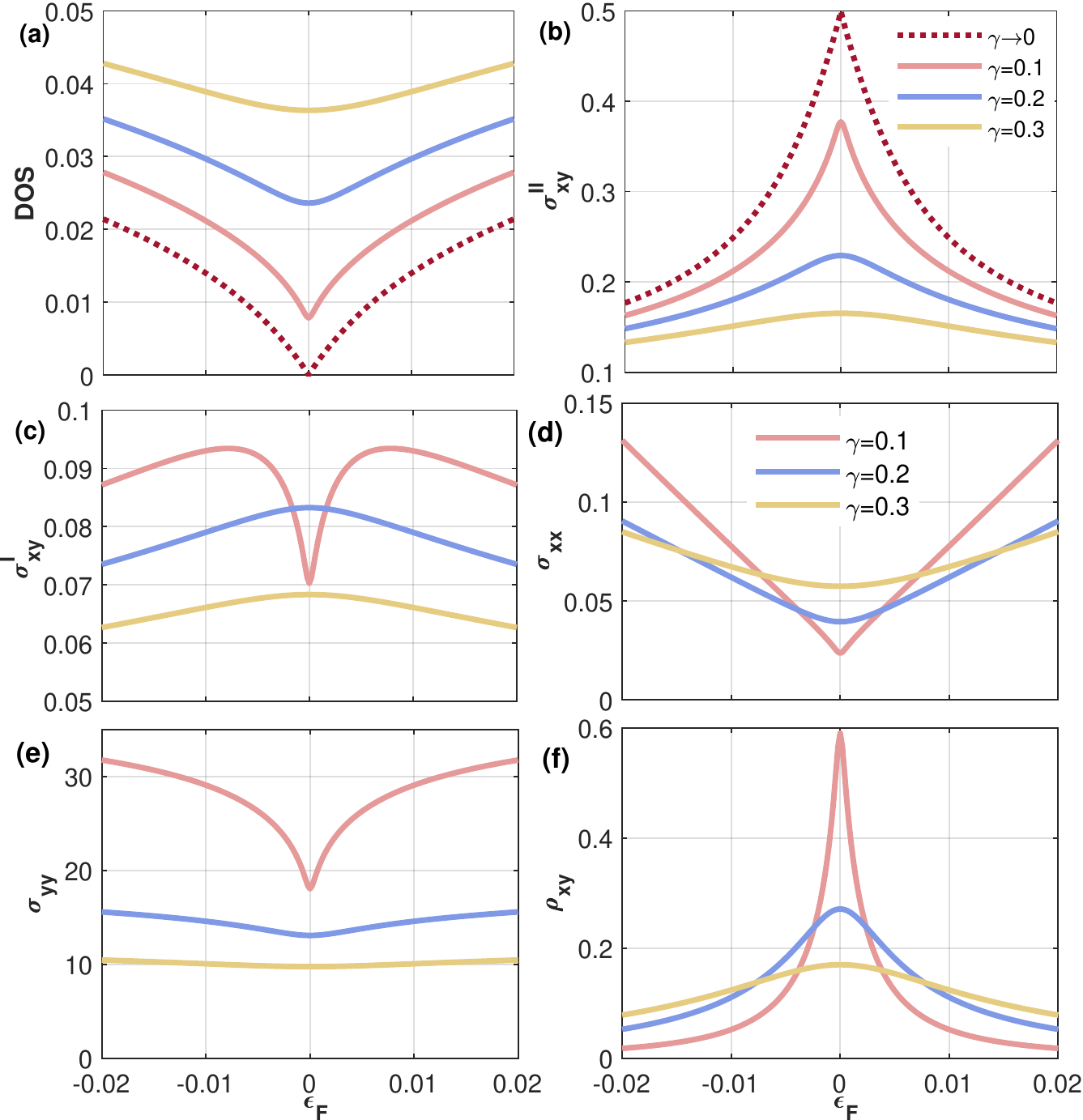}%
	\caption{The disorder-dependent transport properties as a function of Fermi energy $\mathrm{\epsilon_{F}}$. (a) The density of states. (b) The Fermi sea part of Hall conductivity $\sigma^{\mathrm{II}}_{xy}$. (c) The Fermi surface part of Hall conductivity $\sigma^{\mathrm{I}}_{xy}$. In panels (d) and (e), the longitudinal conductivity along the $x$ and $y$ direction $\sigma_{xx}$ and $\sigma_{yy}$, respectively. (f) The anomalous Hall resistivity $\rho_{xy}$. The disorder strengths denoted by $\gamma$ are also inserted, as shown in panels (b) and (d). We set $\lambda=0.1$ in (a)-(f).}
	\label{Fig:Fig1}
\end{figure}


\emph{{\color{magenta} Diffusive transport of topological Dirac semimetals with disorder.}}---For a 2D insulating system, the Hall conductivity remains plateau inside the band gap. In contrast, for the Dirac semimetallic system described by the Hamiltonian Eq.~(\ref{eq:hk}), the half-integer quantization of Hall conductivity $\sigma_{xy}= \frac{e^2}{2h}$ can only exist at Fermi energy $\mathrm{\epsilon_{F}}=0$, which are strongly affected by the presence of disorder. Here, we will investigate the stability of the topological Dirac semimetallic system described by Eq.~(\ref{eq:hk}) against disorder. Let us start by considering the disorder potential given by
\begin{equation}
\begin{aligned}
\hat{V} = \sum_{\boldsymbol{r}}[\mathrm{V_A} (\boldsymbol{r})|\boldsymbol{r},\mathrm{A}\rangle\langle\boldsymbol{r},\mathrm{A}| +
\mathrm{V_B}(\boldsymbol{r}) |\boldsymbol{r},\mathrm{B}\rangle\langle\boldsymbol{r},\mathrm{B}|],
\end{aligned}
\label{eq:Vdis}
\end{equation}
where $\mathrm{V_{A/B}}(\boldsymbol{r}) = \sum_i \mathrm{U_{A/B}}(\boldsymbol{R}_i)\delta(\boldsymbol{r}-\boldsymbol{R}_i)$ and $\mathrm{U_{A/B}}(\boldsymbol{R}_i)$ denote the sublattice uncorrelated impurity potential (see the SM~\cite{SM}) at position $\boldsymbol{R}_i$, which is uniformly distributed in the interval $\mathrm{[-U,U]}$. 
The disorder-averaged retarded (advanced) Green's function $\mathrm{G}^{\mathrm{R}(\mathrm{A})}$ can be expressed as
\begin{equation}
\begin{aligned}
\mathrm{G}^{\mathrm{R}(\mathrm{A})}(\epsilon,\mathbf{k}) &= \frac{1}{\epsilon - \mathrm{h}(\mathbf{k})  -\Sigma^{\mathrm{R}(\mathrm{A})}(\epsilon)},
\end{aligned}
\label{eq:gr}
\end{equation}
where the retarded (advanced) self-energy $\Sigma^{\mathrm{R}(\mathrm{A})}(\epsilon)$ represents the interaction between quasi-particles and disorder. Based on the self-consistent Born approximation (SCBA)~\cite{mahan}, the self-energy function is determined by
\begin{equation}
\begin{aligned}
\Sigma^{\mathrm{R}(\mathrm{A})}(\epsilon) = \gamma\int \frac{d^2\mathbf{k}}{(2\pi)^2}
\mathrm{G}^{\mathrm{R}(\mathrm{A})}[\epsilon-\Sigma^{\mathrm{R}(\mathrm{A})}(\epsilon),\mathbf{k}],
\end{aligned}
\label{eq:selfE}
\end{equation}
where $\gamma = \mathrm{U}^2/3$ denotes the strength of disorder. As discussed in the Supplemental Material~\cite{SM}, the self-energy is diagonal $\Sigma^{\mathrm{R}(\mathrm{A})}(\epsilon)= \Sigma_0\sigma_0$, which does not give any correction to the gap parameter $\delta$ in Eq.~\ref{eq:hking}. Therefore, the Dirac semimetallic state at $\delta= 0$ will not be shifted by the disorder. 
The density of states can be derived from the retarded Green's function Eq.~(\ref{eq:gr}) as  $\rho=-\frac{1}{\pi\gamma}\mathrm{Im}\mathrm{Tr}\mathrm{G}^{\mathrm{R}(\mathrm{A})}$.
As shown in Fig.~\ref{Fig:Fig1}(a), it is found that the disorder induces a finite density of states at the initially gapless point (i.e., $\mathrm{\epsilon_{F}} = 0$), which transforms the semimetal into a diffusive metal~\cite{Mirlin2006}. 

To characterize transport properties in the presence of disorder, we calculate the Hall conductivity based on the Streda-Smrcka decomposition of Kubo formula $\sigma_{xy} = \sigma^{\mathrm{I}}_{xy} + \sigma^{\mathrm{II}}_{xy}$~\cite{Streda-Smrcka1,Streda-Smrcka2}, where
\begin{equation}
\begin{aligned}
\sigma^{\mathrm{I}}_{xy} &= \frac{-e^2\hbar}{4\pi}\int d\epsilon \frac{df(\epsilon)}{d\epsilon}
\mathrm{Tr}[v_{x}(\mathrm{G}^{\mathrm{R}}(\epsilon)-\mathrm{G}^{\mathrm{A}}(\epsilon))v_{y}\mathrm{G}^{\mathrm{A}}(\epsilon)\\
&-v_{x}\mathrm{G}^{\mathrm{R}}(\epsilon)v_{y}(\mathrm{G}^{\mathrm{R}}(\epsilon)-\mathrm{G}^{\mathrm{A}}(\epsilon))],
\end{aligned}
\label{eq:KS-I}
\end{equation}
and
\begin{equation}
\begin{aligned}
\sigma^{\mathrm{II}}_{xy} &= \frac{e^2\hbar}{4\pi}\int d\epsilon f(\epsilon)
\mathrm{Tr}[v_{x}\mathrm{G}^{\mathrm{R}}(\epsilon)v_{y}\frac{d\mathrm{G}^{\mathrm{R}}(\epsilon)}{d\epsilon}
-v_{x}\frac{d\mathrm{G}^{\mathrm{R}}(\epsilon)}{d\epsilon}\times\\
&v_{y}\mathrm{G}^{\mathrm{R}}(\epsilon)-v_{x}\mathrm{G}^{\mathrm{A}}(\epsilon)v_{y}\frac{d\mathrm{G}^{\mathrm{A}}(\epsilon)}{d\epsilon} + v_{x}\frac{d\mathrm{G}^{\mathrm{A}}(\epsilon)}{d\epsilon}v_{y}\mathrm{G}^{\mathrm{A}}(\epsilon)].
\end{aligned}
\label{eq:KS-II}
\end{equation}
The longitudinal conductivity along $\sigma_{\mu\mu}$ can be obtained from the Kubo-Greenwood formula as
\begin{equation}
\begin{aligned}
\sigma_{\mu\mu} &= \frac{-e^2\hbar}{\pi}\int d\epsilon \frac{df(\epsilon)}{d\epsilon}
\mathrm{Tr}[v_{\mu}\mathrm{Im}\mathrm{G}(\epsilon)v_{\mu}\mathrm{Im}\mathrm{G}(\epsilon)],
\end{aligned}
\label{eq:KG}
\end{equation}
where  $\mathrm{Im}\mathrm{G}(\epsilon)=\frac{\mathrm{G}^{\mathrm{R}}(\epsilon)-\mathrm{G}^{\mathrm{A}}(\epsilon)}{2i}$ and $\mathrm{G}^{R(A)}$ are the disorder-averaged retarded (advanced) Green's functions defined by Eq.~(\ref{eq:gr}). 
Based on the effective Hamiltonian Eq.~(\ref{eq:hk}), the velocity operators are $v_{x,y} = \hbar^{-1}\partial_{k_{x,y}}\mathrm{h}_{\mathbf{k}}$.
\noindent{\color{blue}We note that the calculation of the trace ($\mathrm{Tr}$) is preformed over the sublattice space
and momentum space~\cite{SM}}.
The $f(\epsilon)=[e^{(\epsilon-\mathrm{\epsilon_F})/{\mathrm{k_BT}}}+1]^{-1}$ is the Fermi-Dirac distribution with the Fermi energy $\mathrm{\epsilon_{F}}$ and temperature $\mathrm{T}$. Here, we focus on the disorder-dependence of conductivity $\sigma_{\mu\nu}$ and set temperature as $\mathrm{T}=0$. In this case, the function $-\frac{df(\epsilon)}{d\epsilon}$ is reduced to $\delta(\epsilon-\mathrm{\epsilon_{F}})$ at zero temperature. In this case, the Fermi surface term $\sigma^{\mathrm{I}}_{xy}$ and longitudinal conductivity along $\sigma_{\mu\mu}$ are determined by the quantities at the Fermi energy $\mathrm{\epsilon_F}$. By substituting $\mathrm{G}^{\mathrm{R}(\mathrm{A})}(\epsilon,\mathbf{k})$ into Eq.~(\ref{eq:KS-I}), we have
\begin{equation}
\begin{aligned}
\sigma^{\mathrm{I}}_{xy}(\mathrm{\epsilon_{F}}) &= \frac{e^2\hbar}{2\pi}\mathrm{Tr}[v_{x}\mathrm{G}^{\mathrm{R}}(\mathrm{\epsilon_{F}})v_{y}\mathrm{G}^{\mathrm{A}}(\mathrm{\epsilon_{F}})],\\
&=\frac{2e^2}{h}(\mathrm{Z}^{\mathrm{A}}-\mathrm{Z}^{\mathrm{R}})\int \frac{d^2\mathbf{k}}{(2\pi)^2}iv\lambda \mathrm{d}_x(\mathbf{k})\mathrm{g}^{\mathrm{R}}_{\mathbf{k}}(\mathrm{\epsilon_{F}})\mathrm{g}^{\mathrm{A}}_{\mathbf{k}}(\mathrm{\epsilon_{F}}),\\
\end{aligned}
\label{eq:KS2-I}
\end{equation}
where $\mathrm{Z}^{\mathrm{R}(\mathrm{A})} = \mathrm{\epsilon_{F}}-\Sigma^{\mathrm{R}(\mathrm{A})}(\mathrm{\epsilon_{F}})$. We define $\mathrm{g}^{\mathrm{R}(\mathrm{A})}_{\mathbf{k}}=[(\mathrm{Z}^{\mathrm{R}(\mathrm{A})})^2-\mathrm{d}^2_{\mathbf{k}}]^{-1}$ for conciseness. Notice that the other terms in Eq.~(\ref{eq:KS-I}) vanishes $\mathrm{Tr}[v_{x}\mathrm{G}^{\mathrm{R}(\mathrm{A})}v_{y}\mathrm{G}^{\mathrm{R}(\mathrm{A})}]=0$ by setting $\mathrm{A}=\mathrm{R}$ in Eq.~(\ref{eq:KS2-I}). From Eq.~(\ref{eq:KS-II}) we can derive
\begin{equation}
\begin{aligned}
\sigma^{\mathrm{II}}_{xy}(\mathrm{\epsilon_{F}}) &=\frac{2e^2}{h}\int d\epsilon \int \frac{d^2\mathbf{k}}{(2\pi)^2}iv\lambda \mathrm{d}_x(\mathbf{k})[\mathrm{g}^{\mathrm{A}}_{\mathbf{k}}(\epsilon)-\mathrm{g}^{\mathrm{R}}_{\mathbf{k}}(\epsilon)],
\end{aligned}
\label{eq:KS2-II}
\end{equation}
where the integration $\int_{\epsilon}$ in Eq.~(\ref{eq:KS-II}) is restricted to the interval $\int^{\mathrm{\epsilon_{F}}}_{-\infty}$, since the Fermi sea term $\sigma^{\mathrm{II}}_{xy}$ accepts all the contributions blow the Fermi energy $\mathrm{\epsilon_{F}}$.


We first calculate the anomalous Hall conductivity $\sigma^{\mathrm{I}}_{xy}$ and $\sigma^{\mathrm{II}}_{xy}$ in the low-energy region. As shown in Fig.~\ref{Fig:Fig1}(b), the Fermi sea term $\sigma^{\mathrm{II}}_{xy}$ exhibits a peak at $\mathrm{\epsilon_F}=0$ where the half-integer quantized Hall conductivity $\sigma^{\mathrm{II}}_{xy}=\frac{e^2}{2h}$ can be nearly preserved for very weak disorder. However, the increase of disorder can quickly reduce the conductivity and smear the half-integer quantization. On the other hand, as shown in Fig.~\ref{Fig:Fig1}(c), the Fermi surface term $\sigma^{\mathrm{I}}_{xy}$ gives the positive contributions to the anomalous Hall conductivity in the metallic regime but is not enough to compensate the loss in $\sigma^{\mathrm{II}}_{xy}$. Nevertheless, the total quantity $\sigma_{xy} = \sigma^{\mathrm{I}}_{xy} + \sigma^{\mathrm{II}}_{xy}$ is still disorder-dependent and is reduced with increasing disorder.

In Figs.~\ref{Fig:Fig1}(d) and \ref{Fig:Fig1}(e), we can see that the longitudinal conductivities $\sigma_{xx}$ and $\sigma_{yy}$ behave very differently in the diffusive regime due to anisotropy. The conductivity $\sigma_{xx}$ is enhanced as disorder increases. Whereas, the conductivity along the other direction behaves adversely. By inverting the conductivity tensor, the resistivity tensor can be expressed as
\begin{equation}
\begin{aligned}
\rho_{\mu\nu} &= [\sigma_{\mu\nu}]^{-1}
= \frac{1}{\sigma_{xx}\sigma_{yy}+\sigma_{xy}^2}\begin{pmatrix} \sigma_{yy} & -\sigma_{xy} \\ \sigma_{xy} & \sigma_{xx} \end{pmatrix}.
\label{eq:rty}
\end{aligned}
\end{equation}
As shown in Fig.~\ref{Fig:Fig1}(f), the Hall resistivity $\rho_{xy}$ is reduced with increasing disorder. This is different from the insulating phase where the longitudinal conductivity vanishes and $\rho_{xy}$ acts as the inverse of Hall conductivity.

\begin{figure}[htb]
	\centering
	\includegraphics[width=0.49 \textwidth]{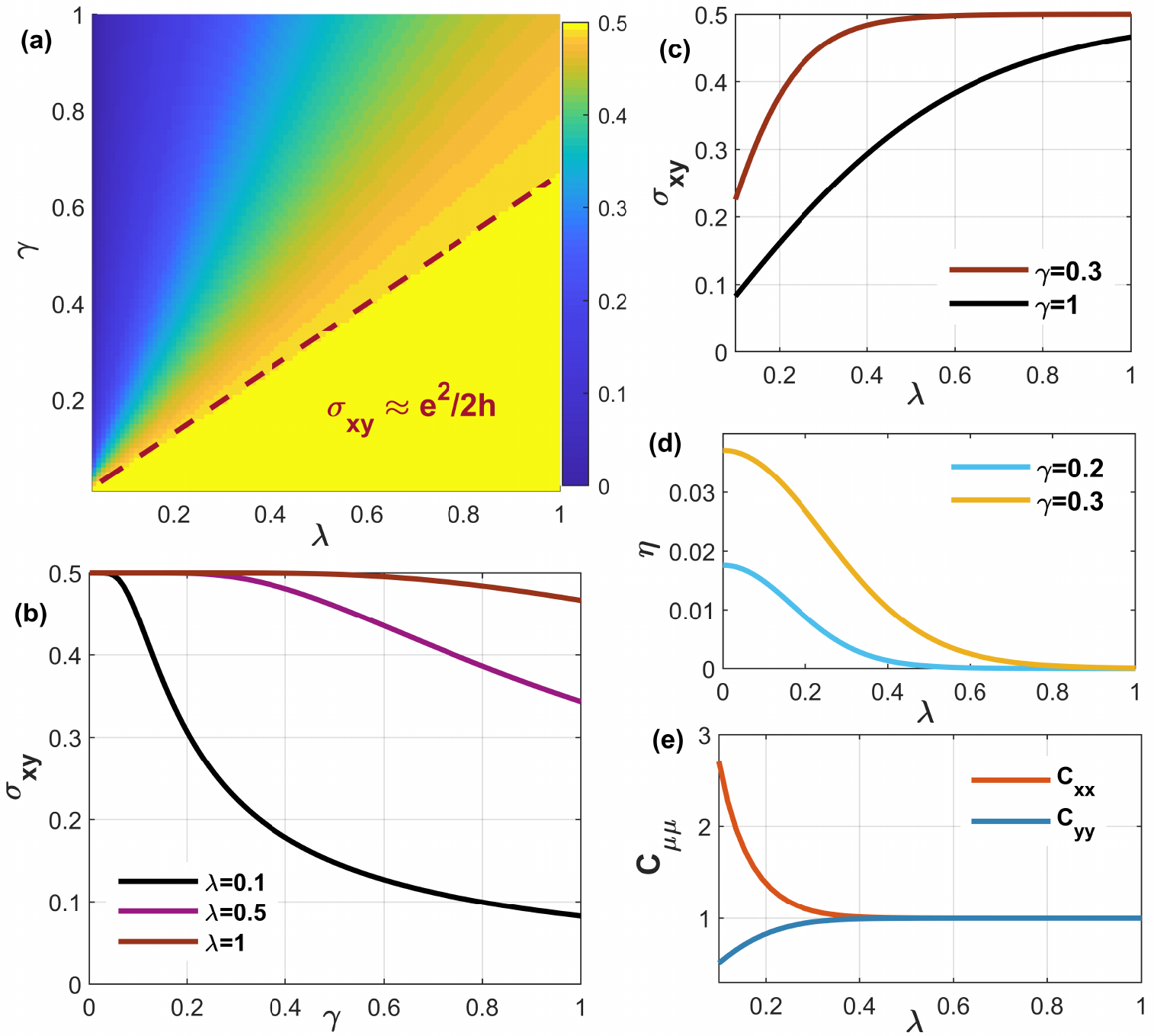}%
	\caption{(a) The phase diagram of the anomalous Hall semimetal depicted by $\sigma_{xy}(\lambda,\gamma)$. (b) The Hall conductivity at $\mathrm{\epsilon_F} = 0$ as a function of damping parameter $\eta$ at fixed $\lambda=0.1$, 0.5, and 1. (c) The Hall conductivity at $\mathrm{\epsilon_F} = 0$ as a function of $\lambda$ at fixed disorder strength $\gamma=0.3$ and 1. (d) The suppressing of the damping parameters with increasing $\lambda$ at fixed $\gamma=0.2$ and 0.3. (e) The normalized longitudinal conductivity $\mathrm{C}_{\mu\mu}=\sigma_{\mu\mu}/\sigma^0_{\mu\mu}$ as a function of $\lambda$.}
	\label{Fig:Fig2}
\end{figure}

\emph{{\color{magenta} Stability of half-integer quantized Hall conductivity against disorder.}}---To explicitly examine the influence of disorder, we focus on the Hall conductivity $\sigma_{xy}$ at $\mathrm{\epsilon_{F}}=0$. Based on Eqs.~(\ref{eq:KS2-I}) and ~(\ref{eq:KS2-II}), we can decompose the disorder-dependent $\sigma_{xy}$ as
\begin{equation}
\begin{aligned}
\sigma_{xy} &= \sigma^{\mathrm{int}}_{xy}+ \delta\sigma_{xy},
\label{eq:dAHC}
\end{aligned}
\end{equation}
where $\delta\sigma_{xy}=\sigma^{\mathrm{I}}_{xy} + \delta\sigma_{xy}^{\mathrm{II}}$, and $\delta\sigma_{xy}^{\mathrm{II}}$ is from the splitting of the Fermi sea term as $\sigma^{\mathrm{II}}_{xy}=\sigma^{\mathrm{int}}_{xy} + \delta\sigma_{xy}^{\mathrm{II}}$.
The intrinsic part can be obtained from Eq.~(\ref{eq:AHC}) as
\begin{equation}
\begin{aligned}
\sigma^{\mathrm{int}}_{xy} = \frac{e^2}{2h}.
\end{aligned}
\label{eq:AHCint}
\end{equation}
The other two terms such as $\delta\sigma_{xy}^{\mathrm{II}}$ and $\sigma^{\mathrm{I}}_{xy}$  are disorder-dependent and given by
\begin{equation}
\begin{aligned}
\delta\sigma_{xy}^{\mathrm{II}} &= \frac{-2e^2}{h}\int^{\infty}_{0} dx F(x)[\frac{x\eta}{x^2+\eta^2}+\arctan{\frac{\eta}{x}}],
\end{aligned}
\label{eq:dKS3-II}
\end{equation}
and
\begin{equation}
\begin{aligned}
\sigma^{\mathrm{I}}_{xy} &= \frac{4e^2}{h}\int^{\infty}_{0} dx F(x)\frac{\eta x^3}{(\eta^2+x^2)^2}.
\end{aligned}
\label{eq:dKS3-I}
\end{equation}
The damping parameter $\eta$ in Eqs.~(\ref{eq:dKS3-II}), (\ref{eq:dKS3-I}) is defined by the imaginary part of self energy as $\mathrm{Im}\Sigma^{\mathrm{R}/\mathrm{A}}= \mp\eta(\lambda,\gamma)$, and we introduce a function $F(x)$ related to the density of the Berry curvature $F(x) = 2\int_{\mathbf{k}} \delta(x-\mathrm{d}_{\mathbf{k}})\Omega(\mathbf{k})$. \noindent{\color{blue}We provide the calculation details of the Hall conductivity and the derivation of Eqs.~(\ref{eq:dKS3-II}), (\ref{eq:dKS3-I}) in the Supplemental Material~\cite{SM}}.

The stability of the half-integer quantized Hall conductivity is controlled by the disorder correction $\delta\sigma_{xy}$ in Eq.~(\ref{eq:dAHC}), which is depicted in a parameter space ($\lambda$, $\gamma$). In Fig.~\ref{Fig:Fig2}(a), we give a phase diagram of $\sigma_{xy}(\lambda,\gamma)$ to characterize the competition between the disorder $\gamma$ and symmetry breaking term $\lambda k_x\sigma_z$. In the strong disorder limit $\eta \rightarrow\infty$, we have $\sigma^{\mathrm{I}}_{xy}=0$ and $\delta\sigma_{xy}^{\mathrm{II}}=-\frac{e^2}{2h}$ which cancels out the intrinsic part of Hall conductivity $\sigma^{\mathrm{int}}_{xy} = \frac{e^2}{2h}$ and annihilates the anomalous Hall effects, i.e., $\sigma_{xy}=0$. On the other hand, one can enhance the anomalous Hall conductivity $\sigma_{xy}$ by increasing $\lambda$ as shown in Fig.~\ref{Fig:Fig2}(c). We find there is a crossover from the diffusive metal into the topological Dirac semimetal where the value of $\sigma_{xy}$ can be  nearly preserved to $\frac{e^2}{2h}$. This process is accompanied by the decrease of the damping parameter $\eta$ as shown in Fig.~\ref{Fig:Fig2}(d). We note that the function $F(x)$ in Eqs.~(\ref{eq:dKS3-II}) and (\ref{eq:dKS3-I}) becomes smooth near $\mathrm{\epsilon_F}=0$ as $\lambda$ increases and can be replaced by its maximum value $F_0=\frac{c_x}{2\pi\lambda^2} + \frac{c_y}{2\pi v^2}$ (at $x=0$). Finally, we complete the integration and the Eq.~(\ref{eq:dAHC}) can be simplified as
\begin{equation}
\begin{aligned}
\sigma_{xy} &\approx \frac{e^2}{h}(\frac{1}{2}-4F_0\eta).
\end{aligned}
\label{eq:dAHCa}
\end{equation}
Generally speaking, the Hall conductivity of topological Dirac semimetals with parity anomaly cannot exactly be half-integer quantized due to the existence of finite damping induced by disorder. However, we can as much as possibly weaken the disorder effects by choosing a larger value of $\lambda$ [see Fig.~\ref{Fig:Fig2}(b)]. In this case, the quantization of $\sigma_{xy}=\frac{e^2}{2h}$ remains very stable in a wide region of disorder.

\begin{figure}[htb]
	\centering
	\includegraphics[width=0.47\textwidth]{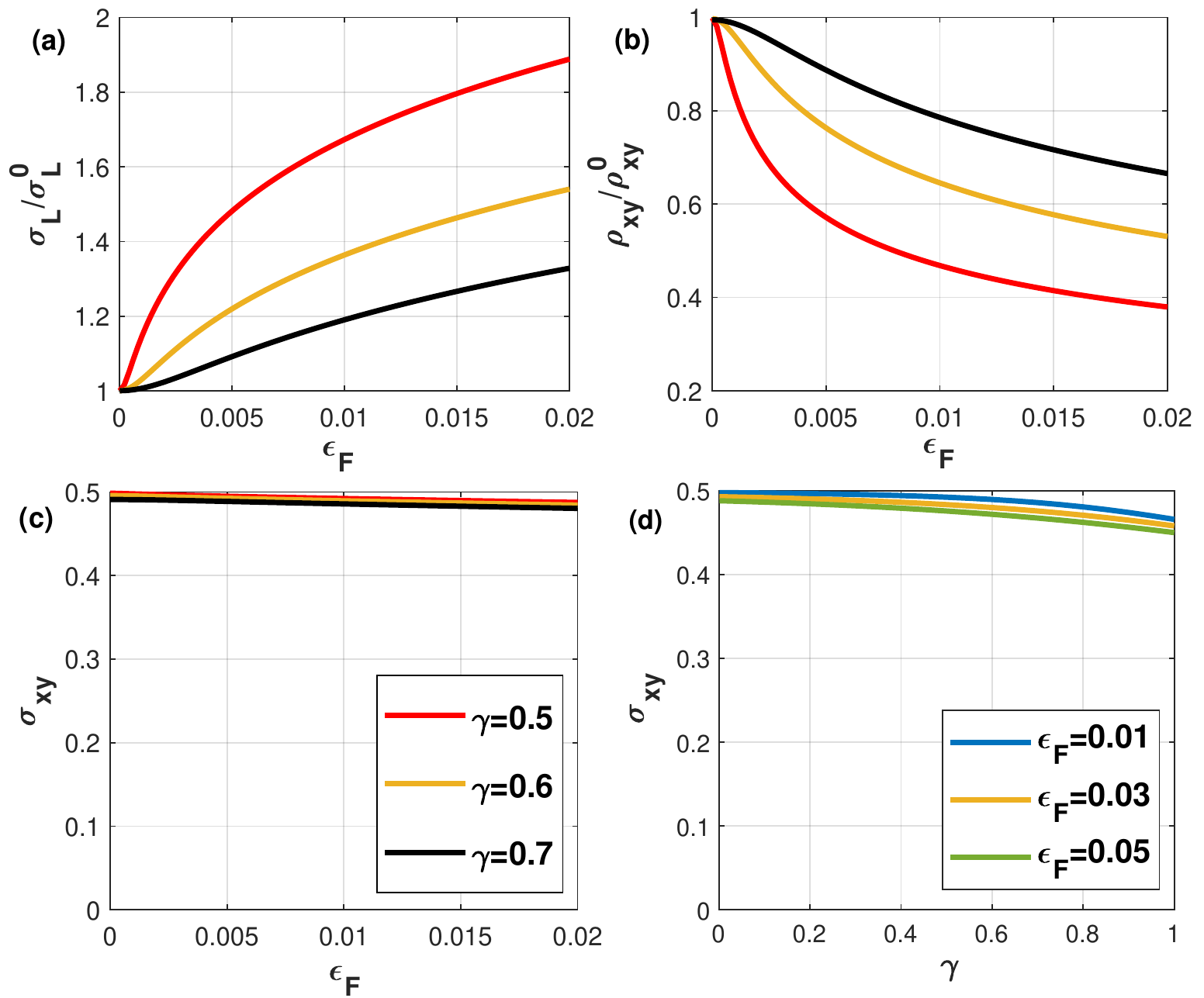}%
	\caption{(a) The longitudinal conductivity defined as $\sigma_L=\sqrt{\sigma_{xx}\sigma_{yy}}$ and is measured by $\sigma^0_{L}=\sqrt{\sigma^0_{xx}\sigma^0_{yy}}=\frac{2e^2}{\pi h}$. (b) The anomalous Hall resistivity $\rho_{xy}$ with $\rho^0_{xy}=\frac{2h}{e^2}(1+\frac{16}{\pi^2})^{-1}$. (c) The Hall conductivity for different disorder strength $\gamma$ as tuning the Fermi level $\epsilon_F$. (d) The disorder-dependence of Hall conductivity at different Fermi energy $\epsilon_F$.
The disorder strengths in (a)-(c) are $\gamma = 0.5, 0.6, 0.7$, and we set $\lambda=1$ in (a)-(d).}
	\label{Fig:Fig3}
\end{figure}

Next, we illustrate the behavior of Hall conductivity $\sigma_{xy}$ by tuning the Fermi level.
As shown in Figs. \ref{Fig:Fig1}(e) and \ref{Fig:Fig1}(f), the finite value of $\sigma_{\mu\mu}$ in the regime of diffusive metals can affect the behavior of anomalous Hall resistivity $\rho_{xy}$, which can also influence the half-integer quantization of $\sigma_{xy}$. With suppressing damping parameter $\eta$, the transport behavior can be substantially changed. To depict this, as shown in Fig.~\ref{Fig:Fig2}(e), we introduce the normalized longitudinal conductivity $\mathrm{C}_{\mu\mu}=\sigma_{\mu\mu}/\sigma^0_{\mu\mu}$ with $\sigma^0_{xx}=\frac{2e^2}{\pi h}\frac{\lambda}{v}$ and $\sigma^0_{yy}=\frac{2e^2}{\pi h}\frac{v}{\lambda}$~\cite{SM}, and we can see that $\mathrm{C}_{\mu\mu}$ approaches to unit with increasing $\lambda$. That is to say, the longitudinal conductivity $\sigma_L = \sqrt{\sigma_{xx}\sigma_{yy}}$ at $\mathrm{\epsilon_F}=0$ approaches to a finite value as universal conductivity constant $\sigma^0_{L}=\frac{2e^2}{\pi h}$, see Fig.~\ref{Fig:Fig3}(a). The similar phenomena can also be observed in
resistivity curve $\rho_{xy}$ [see Fig.~\ref{Fig:Fig3}(b)]. In this case, the Hall resistivity at $\mathrm{\epsilon_F}=0$ approaches to a finite value $\rho^0_{xy}=\frac{2h}{e^2}(1+\frac{16}{\pi^2})^{-1}$ which is not quantized due to the contribution of longitudinal conductivity [see Eq.~(\ref{eq:rty})]. Moreover, the energy dependence of Hall conductivity $\sigma_{xy}$ can be significantly reduced by choosing a larger value of $\lambda$. As shown in Figs.~\ref{Fig:Fig3}(c) and \ref{Fig:Fig3}(d), the nearly half-integer quantization of Hall conductivity can be observed in a relatively wide energy window.

\emph{{\color{magenta}Summary.}}---In summary, we have theoretically propose a strategy to realize the parity anomalous semimetal with a single Dirac cone. Based on the Kubo-Streda formula, we show that the half-integer quantized Hall conductivity, which is a signature of parity anomaly, can be smeared by the damping parameter in the presence of disorder. Through the control of anisotropy of a deformed honeycomb lattice, we can minimize the influence of the disorder and other random fluctuations, maintaining that the Hall conductivity is weakly dependent on the disorder and nearly remains half quantization as $\sigma_{xy}=\frac{e^2}{2h}$. Considering the rapid developments of topological states in the artificial and periodic-driving Floquet systems \cite{Floquet1, Light2022}, we expect that our proposed anisotropic Dirac semimetal with parity anomaly can be realized in experiments. Especially, the recent studies show that the single massless Dirac cone can occur at the phase boundary of external fields induced topological transitions in valley-polarized materials \cite{PhysRevB.105.L081115, PhysRevLett.119.046403,PhysRevLett.112.106802}, indicating that the parity anomalous semimetallic states can be further present in realistic materials and condensed matter systems.



\emph{{\color{magenta}Acknowledgments.}}--- The authors thank Professor Xianggang Wan,  Doctor Bo Fu and Doctor Da-Shuai Ma for useful discussions. This work was supported by the National Natural Science Foundation of China (NSFC, Grants No. 11974062, No. 12222402, No. 12147102, and No. 12074108), and the Beijing National Laboratory for Condensed Matter Physics.

%

\pagebreak
\widetext
\begin{center}
\textbf{Erratum: Robustness of half-integer quantized Hall conductivity against disorder in an anisotropic Dirac semimetal with parity anomaly [Phys. Rev. B 108, L041104 (2023)]}
\end{center}

\makeatletter
\renewcommand{\theequation}{S\arabic{equation}}
\renewcommand{\thefigure}{S\arabic{figure}}
\renewcommand{\bibnumfmt}[1]{[S#1]}
\renewcommand{\citenumfont}[1]{S#1}

We would like to highlight that the concept of the novel topological semimetal phase, known as ``parity anomalous semimetal" which demonstrates a half-quantized Hall effect, was initially introduced in \rm{Ref. [27]} of the main text
and further explored by J.-Y. Zou~\cite{PhysRevB.105.L201106,PhysRevB.107.125153}. The Hamiltonian in Eq. (5) of the main text describes a two dimensional semimetal phase at the critical point $\delta=0$, which can be categorized into the ``parity anomalous semimetal". The three dimensional counterparts known as ``chiral anomalous semimetal" was investigated in \rm{Ref.~\cite{PhysRevB.106.045111}}. Study on Hall conductance and stability of the single Dirac point in the three-dimensional magnetic topological
insulator heterostructure was performed in \rm{Ref.~\cite{PhysRevB.107.125153}}. The Hall conductivity formula $\sigma_{xy} = \frac{e^2}{2h}[1-\mathrm{sgn}(\delta)]$ following the Hamiltonian Eq.(5) is a generic formula which first explicitly presented in the paper~\cite{PhysRevB.81.115407} and was also introduced in the book on topological insulators~\cite{shen2012topological}.

\end{document}